\documentclass[12pt]{article}
\usepackage{amsfonts,amssymb,fullpage}

\title{An Introduction to Noncommutative Geometry}

\author{Joseph C. V\'arilly\\[2\jot]
        Universidad de Costa Rica,\\
        2060 San Jos\'e, Costa Rica}

\date{28 April 2006}

\newcommand{\bC}{\mathbb{C}}            
\newcommand{\bR}{\mathbb{R}}            
\newcommand{\bS}{\mathbb{S}}            
\newcommand{\Dslash}{D\mkern-11.5mu/\,} 
\newcommand{\rSU}{\mathrm{SU}}          
\newcommand{\Spin}{\mathrm{Spin}}       

\newcommand{\chline}[2]{\bigskip\noindent{\bfseries
                         #1\quad #2}\par\smallskip}
\newcommand{\ssline}[2]{\par\indent#1\quad#2}

\begin{document}

\maketitle

\begin{abstract}
The lecture notes of this course at the EMS Summer School on
Noncommutative Geometry and Applications in September, 1997 are now
published by the EMS. Here are the contents, preface and updated
bibliography from the published book.
\end{abstract}

\section*{Contents}

\chline{1\quad Commutative Geometry from the Noncommutative Point of View}

\ssline{1.1}{The Gelfand--Na\u{\i}mark cofunctors}
\ssline{1.2}{The $\Gamma$ functor}
\ssline{1.3}{Hermitian metrics and spin$^c$ structures}
\ssline{1.4}{The Dirac operator and the distance formula}

\chline{2\quad Spectral Triples on the Riemann Sphere}

\ssline{2.1}{Line bundles and the spinor bundle}
\ssline{2.2}{The Dirac operator on the sphere $\bS^2$}
\ssline{2.3}{Spinor harmonics and the spectrum of $\Dslash$}
\ssline{2.4}{Twisted spinor modules}
\ssline{2.5}{A reducible spectral triple}

\chline{3\quad Real Spectral Triples: the Axiomatic Foundation}

\ssline{3.1}{The data set}
\ssline{3.2}{Infinitesimals and dimension}
\ssline{3.3}{The first-order condition}
\ssline{3.4}{Smoothness of the algebra}
\ssline{3.5}{Hochschild cycles and orientation}
\ssline{3.6}{Finiteness of the $K$-cycle}
\ssline{3.7}{Poincar\'e duality and $K$-theory}
\ssline{3.8}{The real structure}

\chline{4\quad Geometries on the Noncommutative Torus}

\ssline{4.1}{Algebras of Weyl operators}
\ssline{4.2}{The algebra of the noncommutative torus}
\ssline{4.3}{The skeleton of the noncommutative torus}
\ssline{4.4}{A family of spin geometries on the torus}

\chline{5\quad The Noncommutative Integral}

\ssline{5.1}{The Dixmier trace on infinitesimals}
\ssline{5.2}{Pseudodifferential operators}
\ssline{5.3}{The Wodzicki residue}
\ssline{5.4}{The trace theorem}
\ssline{5.5}{Integrals and zeta residues}

\chline{6\quad Quantization and the Tangent Groupoid}

\ssline{6.1}{Moyal quantizers and the Moyal deformation}
\ssline{6.2}{Smooth groupoids}
\ssline{6.3}{The tangent groupoid}
\ssline{6.4}{Moyal quantization as a continuity condition}
\ssline{6.5}{The hexagon and the analytical index}
\ssline{6.6}{Quantization and the index theorem}

\chline{7\quad Equivalence of Geometries}

\ssline{7.1}{Unitary equivalence of spin geometries}
\ssline{7.2}{Morita equivalence and connections}
\ssline{7.3}{Vector bundles over noncommutative tori}
\ssline{7.4}{Morita-equivalent toral geometries}
\ssline{7.5}{Gauge potentials}

\chline{8\quad Action Functionals}

\ssline{8.1}{Algebra automorphisms and the metric}
\ssline{8.2}{The fermionic action}
\ssline{8.3}{The spectral action principle}
\ssline{8.4}{Spectral densities and asymptotics}

\chline{9\quad Epilogue: New Directions}

\ssline{9.1}{Noncommutative field theories}
\ssline{9.2}{Isospectral deformations}
\ssline{9.3}{Geometries with quantum group symmetry}
\ssline{9.4}{Other developments}

\section*{Introduction}

This book consists of lecture notes for a course given at the EMS
Summer School on Noncommutative Geometry and Applications, at Monsaraz
and Lisboa, Portugal in September, 1997. These were made available in
preprint form on the ArXiv, as physics/9709045, at that time. In
updating them for publication, I have kept to the original plan, but
have added citations of more recent papers throughout. An extra final
chapter summarizes some of the developments in noncommutative geometry
in the intervening years.

The course sought to address a mixed audience of students and young
researchers, both mathematicians and physicists, and to provide a
gateway to noncommutative geometry, as it then stood. It already
occupied a wide-ranging area of mathematics, and had received some
scrutiny from particle physicists. Shortly thereafter, links to string
theory were found, and its interest for theoretical physicists is now
indisputable.

Many approaches can be taken to introducing noncommutative geometry.
In these lectures, the focus is on the geometry of Riemannian spin
manifolds and their noncommutative cousins, which are `spectral
triples' determined by a suitable generalization of the Dirac
operator. These `spin geometries', which are spectral triples with
certain extra properties, underlie the noncommutative geometry
approach to phenomenological particle models and recent attempts to
place gravity and matter fields on the same geometrical footing.

The first two chapters are devoted to commutative geometry; we set up
the general framework and then compute a simple example, the
two-sphere, in noncommutative terms. The general definition of a spin
geometry is then laid out and exemplified with the noncommutative
torus. Enough details are given so that one can see clearly that
noncommutative geometry is just ordinary geometry, extended by
discarding the commutativity assumption on the coordinate algebra.
Classification up to equivalence is dealt with briefly in
Chapter~7.

Other chapters explore some of the tools of the trade: the
noncommutative integral, the role of quantization, and the spectral
action functional. Physical models are not treated directly (these
were the subject of other lectures at the Summer School), but most of
the mathematical issues needed for their understanding are dealt with
here. The final chapter is a brief overview of the profusion of new
examples and applications of noncommutative spaces and spectral
triples.

I wish to thank several people who contributed in no small way to
assembling these lecture notes. Jos\'e M. Gracia-Bond\'{\i}a gave
decisive help at many points; and Alejandro Rivero provided
constructive criticism. I thank Daniel Kastler, Bruno Iochum, Thomas
Sch\"ucker and the late Daniel Testard for the opportunity to visit
the Centre de Physique Théorique of the CNRS at Marseille, as a
prelude to the Summer School; and Piotr M. Hajac for an invitation to
teach at the University of Warsaw, when I rewrote the notes for
publication. This visit to Katedra Metod Matematycznych Fizyki of UW
was supported by European Commission grant MKTD--CT--2004--509794.

I am grateful for enlightening discussions with Alain Connes, Robert
Coquereaux, Ricardo Estrada, H\'ector Figueroa, Thomas Krajewski,
Giovanni Landi, Fedele Lizzi, Carmelo P\'erez Mart\'{\i}n, William J.
Ugalde and Mark Villarino. Thanks also to Jes\'us Clemente, Stephan
de~Bi\`evre and Markus Walze who provided indispensable references.
Several improvements to the original draft notes were suggested by Eli
Hawkins, Thomas Sch\"ucker and Georges Skandalis. Last but by no means
least, I want to discharge a particular debt of gratitude to Paulo
Almeida for his energy and foresight in organizing the Summer School
in the right place at the right time.

\begin{flushright}
\textit{San Pedro de Montes de Oca}

\textit{April 2006}
\end{flushright}


\subsection*{The Lectures}

The main body of the book is a revised and updated version of the
lectures. A final chapter, reviewing developments up to 2005, has been
added. The book is now published by the European Mathematical Society,
Z\"urich, 2006, as the fourth volume in the \textit{EMS Series of
Lectures in Mathematics}.


\end{document}